\documentclass[12pt,epsf]{article}
\usepackage{amssymb,amsmath}
\usepackage{graphicx}

\newcommand{\be}{\begin{equation}}
\newcommand{\ee}{\end{equation}}
\newcommand{\bea}{\begin{eqnarray}}
\newcommand{\eea}{\end{eqnarray}}
\newcommand{\bear}{\begin{eqnarray}}
\newcommand{\eear}{\end{eqnarray}}
\newcommand{\beas}{\begin{eqnarray*}}

\newcommand{\eeas}{\end{eqnarray*}}
\newcommand{\ba}{\begin{array}}
\newcommand{\ea}{\end{array}}

\newcommand{\la}{\langle}
\newcommand{\ra}{\rangle}

\newcommand{\tr}{\operatorname{tr}}
\newcommand{\pd}[2][1]{\ifnum#1=1 \frac{\partial}{\partial {#2}} \else
  \frac{\partial^#1}{\partial {#2}^{#1}}\fi}
\newcommand{\dpd}[2][1]{\ifnum#1=1 \dfrac{\partial}{\partial {#2}} \else
  \frac{\partial^#1}{\partial {#2}^{#1}}\fi}
\newcommand{\td}[2][1]{\ifnum#1=1 \frac{d}{d{#2}} \else
  \frac{d^#1}{d{#2}^{#1}}\fi}

\newcommand{\nbox}{{\,\lower0.9pt\vbox{\hrule \hbox{\vrule height 0.2 cm \hskip 0.19 cm \vrule height 0.2 cm}\hrule}\,}}

\def\href#1#2{#2}

\begin{document}
\begin{titlepage}
\hfill
\vbox{
    \halign{#\hfil         \cr
           } 
      }  
\vspace*{20mm}
\begin{center}
{\Large \bf Inviolable energy conditions \\ from entanglement inequalities}

\vspace*{15mm}
\vspace*{1mm}
Nima Lashkari, Charles Rabideau, Philippe Sabella-Garnier, Mark Van Raamsdonk
\vspace*{1cm}
\let\thefootnote\relax\footnote{nima, rabideau, psabella, mav @phas.ubc.ca}

{Department of Physics and Astronomy,
University of British Columbia\\
6224 Agricultural Road,
Vancouver, B.C., V6T 1W9, Canada}

\vspace*{1cm}
\end{center}
\begin{abstract}
Via the AdS/CFT correspondence, fundamental constraints on the entanglement structure of quantum systems translate to constraints on spacetime geometries that must be satisfied in any consistent theory of quantum gravity. In this paper, we investigate such constraints arising from strong subadditivity and from the positivity and monotonicity of relative entropy in examples with highly-symmetric spacetimes. Our results may be interpreted as a set of energy conditions restricting the possible form of the stress-energy tensor in consistent theories of Einstein gravity coupled to matter.

\end{abstract}

\end{titlepage}

\vskip 1cm

\section{Introduction}

The AdS/CFT correspondence provides a remarkable connection between quantum gravitational theories and non-gravitational quantum systems \cite{maldacena1997large,aharony2000large}. There are believed to be many examples of the correspondence; indeed, it may be that any consistent quantum gravity theory for asymptotically AdS spacetimes can be used to define a CFT on the boundary spacetime. In this paper, we focus on examples with a classical limit described by Einstein's equations coupled to matter. We seek to derive results that are universally true for all such theories, by translating to gravitational language results that are universally true in all quantum field theories. Specifically, we will translate some basic constraints on the structure of entanglement in quantum systems to derive some fundamental constraints on spacetime geometry that must hold in all consistent theories of Einstein gravity coupled to matter.

Our main tool will be the Ryu-Takayanagi formula (and its covariant generalization due to Hubeny, Rangamani, and Takayanagi)\cite{ryu2006holographic,hubeny2007covariant}.\footnote{A recent proof was given in \cite{Lewkowycz:2013nqa}.} This relates entanglement entropy for spatial regions $A$ in the field theory to the areas of extremal surfaces $\partial A$ in the dual geometry with the same boundary as $A$ (see section 2 for a review). Generally speaking, we can understand this as a mapping
\[
RT : {\bf \cal G} \to {\bf \cal S}
\]
from the set ${\bf \cal G}$ of asymptotically AdS spacetimes with boundary geometry $B$ to the set ${\bf \cal S}$ of maps $S$ from subsets of $B$ to real numbers.\footnote{To avoid divergent quantities, we could define the map $S$ associated with a geometry $M$ such that for subset $B$ of the boundary of $M$, $S_{M}(B)$ is the difference between the area of the extremal surface $\tilde{B}_M$ and the corresponding extremal surface $\tilde{B}_{AdS}$ in pure AdS.}

\begin{figure}
\centering
\includegraphics[width=\textwidth]{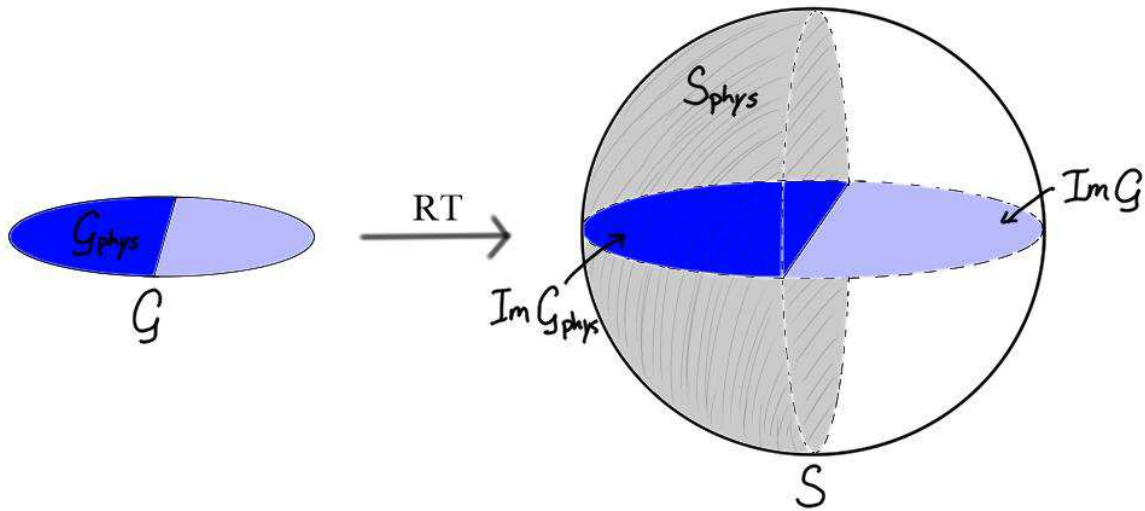}
\caption{Ryu-Takayanagi formula as a map from the space ${\bf \cal G}$ of geometries with boundary $B$ to the space ${\bf \cal S}$ of mappings from subsets of $B$ to real numbers. Mappings in region  ${\bf \cal S}_{phys}$ (shaded) correspond to physically allowed entanglement entropies. Geometries in region ${\bf \cal G}_{phys}$ map into ${\bf \cal S}_{phys}$ while the remaining geometries are unphysical in any consistent theory for which the Ryu-Takayanagi formula holds (plausibly equal to the set of gravity theories with Einstein gravity coupled to matter in the classical limit).}
\label{intropic}
\end{figure}

This mapping is depicted in figure \ref{intropic}. Physically allowed entanglement structures must obey constraints, such as strong subadditivity and positivity/monotonicity of relative entropy, so only a subset ${\cal S}_{phys}$ of maps represented by ${\cal S}$ can represent physically allowed entanglement structures. If a geometry $M \in {\cal G}$ maps to a point outside this subset, we can conclude that such a geometry is not allowed, in any theory for which the Ryu-Takayanagi formula is valid (which we believe to be all consistent gravity theories whose classical limit is Einstein gravity coupled to matter).
Another interesting point is that the space of geometries with boundary $B$ is much smaller than the space of functions on subsets of $B$, so the image of ${\bf \cal G}$ in ${\bf \cal S}$ will be a measure zero subset ${\cal S}_{G}$. This implies that the entanglement structures for quantum field theory states with gravity duals are extremely constrained.

This picture suggests several interesting directions for research:
\begin{itemize}
\item
Characterize the geometries ${\cal G}_{phys}$ that map to physically allowed entanglement entropies ${\cal S}_{phys}$. While some of these geometries may be ruled out by additional constraints not related to entanglement, we can say that any geometry not in ${\cal G}_{phys}$ cannot represent a physical spacetime.
\item
Characterize the constraints on entanglement structure implied by the existence of a holographic dual i.e. understand the subset ${\cal S}_{G}$. Examples include the monogamy of mutual information \cite{Hayden:2011ag}, but there should be much stronger constraints through which the entanglement entropies for most regions are determined in terms of the entanglement entropies for a small subset of regions.
\item
Better understand the inverse mapping from ${\cal S}_{phys}$ to ${\cal G}_{phys}$ to be able to explicitly reconstruct geometries from entanglement entropies.
\end{itemize}
In this paper, we focus on the first direction, though we will have some comments on the second direction in section 6. Many recent papers discuss the third direction, including \cite{Hammersley:2006cp, Bilson:2008ab, Balasubramanian:2013lsa}.

\subsubsection*{Constraining geometry from entanglement}

The question of which geometries give rise to allowed entanglement structures was considered at the level of first order perturbations to pure AdS in \cite{lashkari2014gravitational,faulkner2014gravitation,swingle2014universality} (see also \cite{nozaki2013dynamics,Bhattacharya:2013bna}). Such perturbations correspond to small perturbations of the CFT vacuum state. For these first order CFT perturbations, the entanglement entropy for ball-shaped regions is determined in terms of the expectation value of the stress-energy tensor\footnote{The stress tensor is determined in terms of the entanglement entropy for infinitesimal ball shaped regions, so we can think of the entanglement first law as a constraint determining the entanglement entropies for arbitrary ball-shaped regions from the entanglement entropies for infinitesimal balls.} via the ``first law of entanglement,'' which we review in section 2 below. As shown in \cite{lashkari2014gravitational,faulkner2014gravitation} the gravitational version of this constraint is exactly the linearized Einstein equation. For a discussion of constraints at the second-order in the metric perturbation, see \cite{Banerjee:2014ozp,Banerjee:2014oaa}.

In this paper, we begin to unravel the implications of entanglement constraints on geometries away from this perturbative limit. One might ask whether it is possible to obtain the full non-linear Einstein equations in this way. However, at the classical level, the entanglement quantities tell us only about the dual geometry, so the entanglement constraints will be constraints on the geometries themselves, without reference to any bulk stress-energy tensor. Further, the specific constraints we consider (strong subadditivity of entanglement entropy, and the positivity and monotonicity of relative entropy) take the form of inequalities, so we should expect that the nonlinear constraints also take the form of geometrical inequalities ruling out some geometries as unphysical. This is a natural outcome: since the results must apply to all consistent theories, we cannot expect specific non-linear equations to emerge, but there should be restrictions that apply to the whole class of allowed theories.

In interpreting these geometrical constraints, it is useful to translate them into constraints on the stress-energy tensor {\it assuming} that Einstein's equations hold. This is a very plausible assumption. Indeed, it is possible to argue \cite{swingle2014universality} indirectly using the linearized results that Einstein's equations must be obtained.\footnote{In \cite{swingle2014universality}, it was shown that by considering quantum corrections to the Ryu-Takayanagi formula, the expectation value of the bulk stress-energy tensor comes in as a source for the linearized Einstein equations. Assuming that the source is a generally a local operator, this is enough to see that it must be the stress-energy tensor. It has been argued that the linearized equations together with the stress-energy tensor as a source imply the full non-linear Einstein equations if one demands conservation of the stress-energy tensor in the full theory.} Any geometry provides a solution to Einstein's equations for some stress tensor. Thus, given a geometry that violates the entanglement constraints, we can conclude that no consistent theory of gravity can produce the associated stress tensor. Expressed in this way, the constraints from entanglement inequalities can be thought of as certain ``energy conditions.''

We will see that some of the conditions we obtain are closely related to some of the standard energy conditions used in classical general relativity. However, we emphasize that while these standard conditions (such as the weak and null-energy conditions) are simply plausible assumptions on the properties of matter, the conditions we derive follow from fundamental principles of quantum mechanics (assuming the Ryu-Takayanagi formula holds) and cannot be violated.

\subsubsection*{Summary of Results}

In this paper, we take a few modest steps towards understanding the general constraints on non-linear gravity due to entanglement inequalities, investigating these constraints in the case of highly symmetric spacetimes. Specifically, we determine constraints on static, translationally invariant spacetimes in 2+1 dimensions, and static, spherically-symmetric spacetimes in general dimensions. We find the following main results:
\begin{itemize}
\item
For spacetimes dual to the vacuum states of 1+1 dimensional Lorentz-invariant field theories flowing between two CFT fixed points, the constraints due to strong-subadditivity are satisfied if and only if the spacetime satisfies a set of averaged null energy conditions
\[
\int_\gamma ds T_{\mu \nu} u^{\mu} u^{\nu} \ge 0
\]
where $\gamma$ is an arbitrary spatial geodesic and $u^\mu$ is a null vector generating a light-sheet of $\gamma$ defined such that translation by $u^\mu$ produces an equal change in the spatial scale factor at all points (section 3).
\item
For static translation-invariant spacetimes dual to excited states of 1+1 dimensional CFTs, we show that the monotonicity of relative entropy implies that the minimum scale factor reached by an RT surface for spatial interval is always less than the scale factor reached by the corresponding RT surface in the geometry for the thermal state with the same stress-energy tensor (section 4).
\item
For these spacetimes, we find that asymptotically, the positivity of relative entropy is exactly equivalent to the statement that observers near the boundary moving at arbitrary velocities in the field theory direction cannot observe negative energy. That is, we get a subset of the weak energy condition $T_{\mu \nu} u^\mu u^\nu \ge 0$ where $u^\mu$ is an arbitrary timelike vector with no component in the radial direction.
\item
For static spherically symmetric asymptotically AdS spacetimes, the positivity of relative entropy implies that the area of a surface bisecting the spacetime symmetrically is bounded by the mass of the spacetime. For four-dimensional gravity, the specific result is (section 5)
\[
\Delta A \le 2 \pi G_N M \ell_{AdS} \; .
\]
\end{itemize}
We offer a few concluding remarks in section 6.

Previous connections between energy conditions and entanglement inequalities appeared in \cite{Myers:2012ed, Callan:2012ip, Wall:2012uf, Caceres:2013dma} who noted that the null energy condition is sufficient to prove certain entanglement inequalities holographically. The use of relative entropy in holography was pioneered in \cite{blanco2013relative} and applied to derive gravitational constraints at the perturbative level in $\cite{Banerjee:2014ozp, Banerjee:2014oaa}$.

Note: While this manuscript was in preparation, the paper \cite{Lin:2014hva} appeared, which overlaps with the results in section 4.2.

\section{Background}

\subsection{Entanglement inequalities}

In this section, we review various entanglement inequalities that should place constraints on possible dual spacetimes via the holographic entanglement entropy formula.\footnote{See, for example \cite{nielsen2010quantum}, for a more complete discussion of entanglement inequalities.}

\subsubsection*{Strong subadditivity}

To begin, we recall that the entanglement entropy $S(A)$ for a subsystem $A$ of a quantum system is defined as $S(A) = -\tr(\rho_A \log(\rho_A))$, where $\rho_A$ is the reduced density matrix for the subsystem.

The strong subadditivity of entanglement entropy states that for any three disjoint subsystems $A$, $B$, and $C$,
\be
\label{SSA}
S(A \cup B) + S(B \cup C) \ge S(B) + S(A \cup B \cup C).
\ee
Considering only spatial regions of a constant-time slice in a time-invariant state corresponding to a static dual geometry, this constraint places no constraints on the dual geometry, as shown in \cite{headrick2007holographic}. However, in the time-dependent cases, or for regions of a time-slice that do not respect the symmetry, this inequality gives non-trivial constraints, as we will see below.

\begin{figure}
\centering
\includegraphics[width=0.6\textwidth]{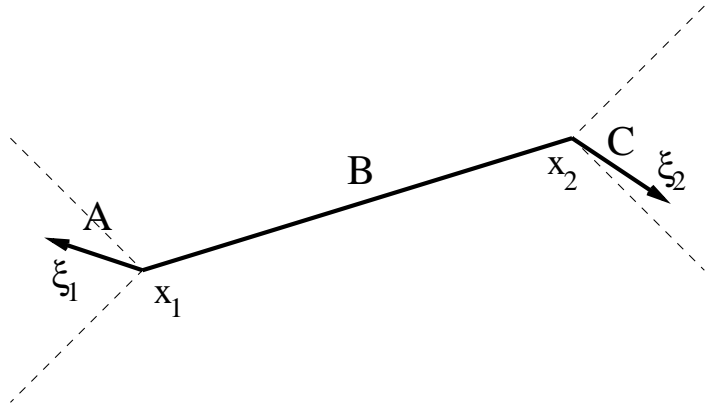}
\caption{Spacelike intervals for strong subadditivity.}
\label{figSSA}
\end{figure}

For our analysis below, we will be interested in applying the constraints of strong subadditivity in the case of 1+1 dimensional field theories. Entanglement entropy is the same for any spacelike regions with the same domain of dependence, so for any connected spacelike region $A$, entanglement entropy is a function of the two endpoints of the region. We write $S(x_1,x_2)$ to denote the entanglement entropy of the interval $[x_1,x_2]$ (or any spacelike region with the same domain of dependence). We focus on the case where A, B, and C in (\ref{SSA}) are adjacent spacelike intervals, as shown in figure \ref{figSSA}.

We note first that the full set of strong subadditivity constraints for adjacent intervals follow from the constraints in the case where the intervals $A$ and $C$ are infinitesimal. For suppose the strong-subadditivity constraint is true for regions $A$, $B$, and $C$ with the proper length of $A$ and $C$ less than $L_{max}$. Then we can show that the constraint holds for intervals with $A$ and $C$ less than $2 L_{max}$, and so forth. For example, if $A$, $B$, $C_1$ and $C_2$ are adjacent intervals with $C_1$ and $C_2$ having proper length less than $L_{max}$, we have
\beas
S(A \cup B) + S(B \cup C_1) &\ge& S(A \cup B \cup C_1) + S(B) \cr
S(A \cup B \cup C_1) + S(B \cup C_1 \cup C_2) &\ge& S(A \cup B \cup C_1 \cup C_2) + S(B \cup C_1)
\eeas
Adding these, we find
\[
S(A \cup B) + S(B \cup \left\{C_1 \cup C_2 \right\}) \ge S(A \cup B \cup \left\{C_1 \cup C_2 \right\}) + S(B) \; .
\]
In this way, we can combine two strong subadditivity constraints for which the rightmost interval has length smaller than $L_{max}$ to obtain a constraint where the rightmost interval is any interval with length less than $2 L_{max}$.\footnote{Essentially the same argument works in general dimensions to show that the full set of strong subadditivity constraints are implied by considering the constraint (\ref{SSA}) where $B$ is an arbitrary region and where $A$ and $C$ are taken to be infinitesimal.}

Now, consider the strong subadditivity constraint where $B$ is the interval $[x_1,x_2]$ while $A$ and $C$ are the intervals $[x_1+\epsilon \xi_1,x_1]$ and $[x_2,x_2 + \delta \xi_2]$, as shown in figure \ref{figSSA}. In this case, the constraint (\ref{SSA}) gives
\beas
&& S([x_1+\epsilon \xi_1,x_1] \cup [x_1,x_2]) + S([x_1,x_2] \cup [x_2 + \delta \xi_2]) \cr && \hskip 2 in \ge S([x_1+\epsilon \xi_1,x_1] \cup [x_1,x_2] \cup [x_2 + \delta \xi_2]) + S([x_1,x_2]) \cr
\implies && S(x_1+\epsilon \xi_1,x_2) + S(x_1,x_2 + \delta \xi_2) - S(x_1+\epsilon \xi_1,x_2 + \delta \xi_2) - S(x_1,x_2) \ge 0
\eeas
Expanding to first order in both $\delta$ and $\epsilon$, this gives
\[
\xi_1^\alpha \xi_2^\beta \partial^1_\alpha \partial^2_\beta S(x_1,x_2) \le 0 \; .
\]
Since this constraint is linear in the spacelike vectors $\xi_1$ and $\xi_2$, it is sufficient to require that the constraint be satisfied in the lightlike limit of $\xi_1$ and $\xi_2$, i.e. when  $\xi_1$ and $\xi_2$ lie along the dotted lines in figure \ref{figSSA}. Thus, a minimal set of strong subadditivity constraints that imply all constraints for connected regions is
\[
\partial^1_+ \partial^2_+ S(x_1,x_2) \le 0 \qquad
\partial^1_+ \partial^2_- S(x_1,x_2) \le 0 \qquad
\partial^1_- \partial^2_+ S(x_1,x_2) \le 0 \qquad
\partial^1_- \partial^2_- S(x_1,x_2) \le 0\; .
\]

In the special case of states invariant under spacetime translations, the entanglement entropy for an interval can only depend on the difference between the endpoints so $S(x_1,x_2) = S(x_2 - x_1)$. In this case, the basic constraints may be written as\footnote{Similar constraints were noted in \cite{Bhattacharya:2014vja}, which appeared while the current version of this paper was in preparation.}
\be
\partial_+ \partial_+ S(x) \le 0 \qquad
\partial_- \partial_- S(x) \le 0 \qquad
\partial_+ \partial_- S(x) \le 0 \qquad
\partial_- \partial_+ S(x) \le 0 ; .
\label{SSAti}
\ee
Only the latter two constraints here are saturated for the vacuum state, so we expect these will provide more useful constraints.

Finally, in the case of a Lorentz-invariant state, the entanglement entropy can depend only on the proper length of the interval, so is described by a single function $S(R)$. In this case, the constraints reduce to
\be
{d^2 S \over d R^2} \pm {1 \over R} {d S \over dR} \le 0 \; ,
\label{SSAli}
\ee
where the first two constraints in (\ref{SSAti}) give the $-$ sign and the latter two give the $+$ sign. In particular, the constraint with the $+$ sign (which is saturated for vacuum states) is equivalent to
\[
c'(R) \le 0 \qquad \qquad c(R) \equiv R {d S \over d R} \; .
\]
This was shown by Casini and Huerta \cite{Casini:2004bw} in their proof of the c-theorem using strong-subadditivity.

\subsubsection*{Positivity and Monotonicity of Relative Entropy}

A very general class of constraints on the entanglement structure of a quantum system are related to {\it relative entropy}. This gives a measure of distinguishability of a density matrix $\rho$ to a reference state $\sigma$, defined as
\[
S(\rho || \sigma) = \tr(\rho \log \rho) - \tr(\rho \log \sigma) \; .
\]
Relative entropy is always positive, increasing from zero for identical states $\rho$ and $\sigma$ to infinity for orthogonal states. Furthermore, for reduced density matrices $\rho_A$ and $\sigma_A$ obtained by a partial trace operation from $\rho$ and $\sigma$, we have
\bea
S(\rho_A\|\sigma_A)\leq S(\rho\|\sigma).
\eea
This decrease in $\rho$ under restriction to a subsystem is known as the monotonicity of relative entropy, or the data processing inequality \cite{nielsen2010quantum}.

It is useful to define the modular Hamiltonian associated with the reference state as $H_\sigma = - \log(\sigma)$, in analogy with thermodynamics. Using this, and the definition $S(\rho) = -\tr( \rho \log( \rho))$ for entanglement entropy, we can rewrite the expression for relative entropy as
\bea\label{deltaF}
S(\rho || \sigma) &=& \tr(\rho \log \rho) - \tr(\sigma \log \sigma)  +  \tr(\sigma \log \sigma) - \tr(\rho \log \sigma) \cr
&=& \langle - \log \sigma \rangle_\rho - \langle - \log \sigma \rangle_\sigma  - S(\rho) + S(\sigma) \cr
&=& \Delta  \langle H_\sigma \rangle - \Delta S.
\label{reldef}
\eea

For nearby states, $\rho-\sigma=\epsilon X$ with $\epsilon\ll 1$ and $X$ an arbitrary traceless Hermitian operator, one can expand relative entropy in powers of $\epsilon$. To the first order in $\epsilon$ relative entropy vanishes. This is typically referred to as the first law of entanglement since it implies $\delta \la H_\sigma\ra=\delta S$. The expression at second order in $\epsilon$ is known as {\it Fisher  information}, and is discussed in detail in section 3.

The rewriting in (\ref{deltaF}) becomes useful in cases where we can compute the modular Hamiltonian $H_\sigma$. Generally this is possible when the reference state is thermal with respect to some Hamiltonian. For example, the density matrix for a half-space in the vacuum state of a Lorentz-invariant field theory on Minkowski space is thermal with respect to the Rindler Hamiltonian (boost generator), so we have $H_{mod} = c \int d^d x x T_{00}$. The cases we consider below can all be obtained by conformal transformations from this example \cite{casini2011towards,blanco2013relative}.

For a ball shaped region in the vacuum state of a CFT on $R^{d,1}$, we have \cite{blanco2013relative}
\be
\label{HmodBall}
H_B = 2 \pi \int_{|x|<R} d^d x \frac{R^2 - |x|^2}{2R}  T^{\text{CFT}}_{00} \; .
\ee
For a ball-shaped region in the vacuum state of a CFT on a sphere, we have
\be
\label{Hmodsphere}
H_B = 2 \pi \int_B d^d x {\cos(\theta) - \cos(\theta_0) \over \sin(\theta_0)} T_{00} \; .
\ee

In the special case of 1+1 dimensional CFTs the modular Hamiltonian can also be calculated for thermal states. For a spatial interval $[-R,R]$ in an unboosted thermal state with temperature $T = \beta^{-1}$, the modular Hamiltonian is
\be
\label{HmodTherm}
H_B = {2 \beta \over \sinh \left({2 \pi R  \over \beta}\right)} \int_{-R}^R d x \sinh\left({\pi (R-x) \over \beta} \right) \sinh\left({\pi (R+x) \over \beta} \right) T_{00}(x) \; ,
\ee
We can also obtain the expression for the modular Hamiltonian of an interval in a boosted thermal state. This is derived in appendix \ref{AppC}.

\subsubsection*{Optimal relative entropy constraints for a family of reference states}

In various situations, we may have a family of reference states $\sigma_\alpha$ depending on parameters $\alpha_i$ (e.g. temperature), and we would like to find the strongest relative entropy constraint coming from this family. We will assume that the modular Hamiltonians for these reference states take the form of an integral over linear combination of local operators with $\alpha$-dependent coefficients,
\be
\label{Hform}
H_\alpha = \int d^d x f_n(x, \alpha) {\cal O}_n(x) \; .
\ee
According to the entanglement first law, under first order variation of the reference state $\sigma_\alpha$, the entanglement entropy of this state changes as
\[
\delta S_\alpha = \int d^d x f_n(x, \alpha) \delta \langle {\cal O}_n(x) \rangle_\alpha \; .
\]
Here the right side corresponds to the variation in the expectation value of the modular Hamiltonian for the reference state under a variation of the state (while keeping the modular Hamiltonian fixed).
Using this result and the definition (\ref{reldef}), we have
\bea
\delta S(\rho || \sigma^\alpha) &=&  \delta \left\{ \langle H_\beta \rangle_{\rho} - \langle H_\beta \rangle_{\sigma^\beta} - S(\rho) + S(\sigma^\beta) \right\} \cr
&=&   \int d^d x \delta f_n(x, \alpha) \left[\langle {\cal O}_n(x) \rangle_{\rho} - \langle {\cal O}_n(x) \rangle_{\sigma_\alpha} \right] \; .
\label{delSdelalpha}
\eea
Thus, the relative entropy will be extremized with respect to parameters $\alpha_i$ if we can choose a reference state such that
\be
 \int d^d x {\partial f_n(x, \alpha) \over \partial \alpha_i} \left[\langle {\cal O}_n(x) \rangle_{\rho} - \langle {\cal O}_n(x) \rangle_{\sigma_\alpha} \right] =0  \; .
\label{delSdelalpha1}
\ee
In the special case where the initial state and reference states are translation invariant, this becomes
\be
{\partial I_n(\alpha) \over \partial \alpha_i} \left[\langle {\cal O}_n \rangle_{\rho} - \langle {\cal O}_n \rangle_{\sigma_\alpha} \right] =0  \; ,
\label{delSdelalpha2}
\ee
where
\[
I_n(\alpha) = \int d^d x  f_n(x, \alpha) \; .
\]
so we see that an extremum will be obtained if we can choose a reference state with the same expectation value as our state for each of the operators,
\be
\label{matchO}
\langle {\cal O}_n \rangle_{\rho} = \langle {\cal O}_n \rangle_{\sigma_\alpha}
\ee
The same state will also be provide an extremum for the monotonicity constraint, since if $R$ parameterizes a region whose size increases with $R$,
\[
 {d \over d \alpha_i} {d \over d R}  S(\rho || \sigma^\alpha) = {\partial^2 I_n(\alpha,R) \over \partial R \partial \alpha_i} \left[\langle {\cal O}_n \rangle_{\rho} - \langle {\cal O}_n \rangle_{\sigma_\alpha} \right] =0 \; .
\]
Thus the reference state $\sigma^{\alpha^*}$ whose operator expectation values match the state $\rho$ will also give the minimum $d S(\rho || \sigma^\alpha) / d R$ (and thus the strongest monotonicity constraint), assuming that the extremum is a minimum.\footnote{In practice, we should also check whether other extrema exist, and check the boundary of the parameter space. However, since the relative entropy provides a measure of how close our state is to the reference state, it is plausible that the relative entropy is minimized by matching the expectation values of operators. For the cases below, we have explicitly checked that this is the case using the explicit form of the modular Hamiltonian.}

The matching of operator expectation values and the form (\ref{Hform}) of the Hamiltonian implies that $\Delta \langle H_{\alpha^*} \rangle =0$, so in this case, the constraint from positivity and monotonicity of relative entropy are simply that
\be
\label{optifinal}
S(\rho_R) - S(\sigma_R^{\alpha^*}) \le 0 \qquad \qquad {d \over d R} (S(\rho_A) - S(\sigma_A^{\alpha^*})) \le 0.
\ee

\subsection{Holographic formulae for entanglement entropy}

In this paper, we consider general theories of gravity dual to holographic QFTs such that the leading order (in the $1/N$ expansion) entanglement entropy for spatial regions of the field theory is computed by the Ryu-Takayanagi formula, or its covariant generalization. This states that the entanglement entropy of a region $A$ is given by
\[
S(A) = {{\rm Area}(\tilde{A}) \over 4 G_N} \; ,
\]
where $\tilde{A}$ is the extremal surface in the dual geometry with $\partial \tilde{A} = \partial A$ (i.e. such that $A$ and $\partial A$ have the same boundary). The surface $\tilde{A}$ is also required to be homologous to $A$, and in cases where multiple extremal surfaces exist, it is the extremal surface with least area.

The Ryu-Takayanagi formula receives quantum corrections from the entanglement entropy of bulk quantum fields, but we consider only the classical limit in this paper. We note also that for theories of gravity with higher powers of curvature or higher derivatives, the entropy is computed using a more complicated functional than area. However, we restrict attention in this paper to theories for which the gravitational sector is Einstein gravity.

\subsection{Energy conditions}

To end this section, we briefly review a few of the standard energy conditions discussed in the gravitational literature. These are statements about the stress-energy tensor that are taken to be plausibly true, but which are generally not derived from any underlying quantum theory.\footnote{See \cite{Parikh:2014mja} for a recent argument for the null-energy condition based on perturbative string theory.} The {\it weak energy condition} states that the energy density in any frame of reference must be non-negative. Specifically, if $u^\mu$ is a timelike vector, then
\[
T_{\mu \nu} u^\mu u^\nu \ge 0 \; .
\]
The {\it null energy condition} takes the same form, but with $u$ is taken to be a null vector. This is implied by the weak energy condition.

Various authors have also considered averaged energy conditions, in which the conditions are only required to hold when averaged over some geodesic or spatial region. This is the type of contraint that we will find below.

\section{Constraints on spacetimes dual to Lorentz-invariant 1+1D field theories}

In this section, we consider Lorentz-invariant holographic two-dimensional field theories that flow from some CFT in the UV to another CFT in the IR. For such theories, the vacuum state is dual to a spacetime of the form\footnote{In special cases, there may be additional compact directions in the dual spacetime. In these cases, we consider the KK-modes of the metric and other fields as part of the matter sector.}
\be
\label{Lorentz}
ds^2 = {F^2(r) \over r^2} dr^2 + r^2 (-dt^2 + dx^2) \; ,
\ee
where $F(r)$ approaches constants both at $r=0$ and at $r= \infty$ (giving AdS geometries corresponding to the IR and UV fixed points).\footnote{This choice of coordinates assumes that the spatial scale factor is monotonic in the radial direction. At the end of this section, we comment on the case where this doesn't hold.} We would like to understand the constraints on the function $F(r)$ that arise from entanglement inequalities in the CFT. Specifically, we consider the constraints arising from strong subadditivity.

For any spacelike interval, Lorentz-invariance implies that the entanglement entropy depends only on the proper length of the interval, so entanglement entropy for connected regions is captured by a single function $S(R)$. As we reviewed in section 2, Casini and Huerta have shown \cite{Casini:2004bw} starting from strong subadditivity that the function $c(R) = d S / d (\ln(R)) = R d S / d R $ obeys $c'(R) \le 0$. The function $c(R)$ therefore decreases monotonically for increasing $R$, which leads immediately to the Zamolodchikov c-theorem, since $c(R)$ reduces to the UV and IR central charge for small and large $R$ respectively.

The holographic version of the statement $c'(R) \le 0$ was obtained previously in \cite{Myers:2012ed}, but we review the calculation here since we will be generalizing this in the next section. Using the Ryu-Takayanagi formula, the entanglement entropy for an interval of length $R$ in the geometry (\ref{Lorentz}) is obtained by the minimum of the action
\be
\label{RTac}
S = \int d \lambda \sqrt{{F^2(r) \over r^2} \left({dr \over d \lambda} \right)^2 + r^2  \left({dx \over d \lambda}\right)^2} \;
\ee
with boundary conditions $(r(\lambda_i), x(\lambda_i)) = (r_{max},0)$ and $(r(\lambda_f), x(\lambda_f)) = (r_{max},R)$, where $r_{max}$ is a regulator that we will take to infinity. In appendix B, we derive a general formula for the variation of the entanglement entropy under a variation in the endpoints of the interval for translation-invariant geometries. For the case of variations in the size of spatial interval, the result (derived previously in \cite{Myers:2012ed}) is that $dS \over d R$ equals the minimum spatial scale factor reached by the RT surface. Thus, for our choice of coordinates,
\be
\label{dSdR}
{dS \over d R} = r_0  \qquad \qquad c(R) = r_0 R\; .
\ee

To find an explicit relation between $r_0$  and $R$ (and check that $r_0$ has a well-defined limit as we remove the regulator), we note that the equation for curves $x(r)$ extremizing the action (\ref{RTac}) is
\[
{d \over dr} \left( {r^2 {d x \over dr} \over \sqrt{{F(r)^2 \over r^2} + r^2 \left({ dx \over dr}\right)^2}}\right) = 0 \; .
\]
In terms of the $r_0$, the value of $r$ where $dr/dx$ vanishes, we have
\be
\label{curveeq}
\left({ dx \over dr}\right)^2 = { F^2(r) \over r^4 \left({r^2 \over r^2_0} - 1 \right)} \; .
\ee
Thus, we obtain
\bea
\label{defw0}
R &=& 2 \int_{r_0}^\infty dr {F(r) \over r^2} {1 \over \sqrt{{r^2 \over r_0^2} - 1}}  \cr
&=&  2 \int_{1}^\infty dx {F(r_0 x) \over r_0 x^2 \sqrt{x^2 - 1}}.
\eea

We can now translate the strong-subadditivity condition $c'(R) \le 0$ to a convenient bulk expression. Starting from the relation (\ref{dSdR}), we have that
\be
\label{derC}
{d \over dR} c(R) =  {dr_0 \over dR} {d \over dr_0} (R r_0) =  {d^2 S \over dR^2}  \int_1^\infty dx {F'(r x) \over x \sqrt{x^2 - 1}}
\ee
Strong subadditivity implies that\footnote{To see this, apply the strong subadditivity constraint (\ref{SSA}) to the case where B is an interval of length R and A and C are intervals of length $\delta R$ to the left and right. Then strong subadditivity implies that  $2S(R + \delta R)-S(R)-S(R + 2\delta R) \ge 0$ which gives $S''(R) \le 0$ in the limit $\delta R \to 0$. Holographically, this implies that Ryu-Takayanagi surfaces for larger intervals must penetrate deeper into the bulk.}
\be
\label{d2S}
{d r_0  \over  d R} = {d^2 S \over  dR^2} \le 0 \; ,
\ee
so we have finally that ${d \over dR} c(R) \le 0$ is equivalent to the condition on $F(r)$ that
\be
\label{RGcondition}
\int_{r_0}^\infty dr {F'(r) \over r \sqrt{{r^2 \over r_0^2}- 1}} \ge 0
\ee
for every $r_0$. This result was derived originally in \cite{Myers:2012ed}.

\subsection{An averaged null energy condition}

We will now show that the condition (\ref{RGcondition}) can be interpreted as a particular averaged null energy condition in this geometry. We first define a null vector field covariantly by the conditions that $u \cdot u = 0$, $u \cdot \xi = 0$, and $u^\mu \partial_\mu r = 1$, where $\xi$ is the Killing vector corresponding to spatial translations along the field theory direction, and the scale factor $r$ can be defined as $r = \sqrt{\xi \cdot \xi}$. In our coordinates, we have $(u^t, u^r, u^x) = (F(r)/r^2,1,0)$. Physically, this null vector field is defined so that translation by the vector field produces the same (additive) change in the scale factor everywhere.

Defining $T_{\mu \nu}$ to be the stress tensor giving rise to the geometry (\ref{Lorentz}) via Einstein's equations, we find that
\[
T_{\mu \nu} u^\mu u^\nu \propto {F'(r) \over r F(r)} \; ,
\]
where we have used that the Einstein tensor in our geometry is
\beas
G_{rr} &=& {1 \over r^2} \cr
G_{tt} = - G_{xx} &=& {r^3 \over F(r)^3} F'(r) - {r^2 \over F(r)^2} \; .
\eeas
From (\ref{Lorentz}) and (\ref{curveeq}) the distance element along an RT curve $B$ with minimal radial coordinate $r_0$ is given by
\[
ds = {dr F(r) \over r_0 \sqrt{{r^2 \over r_0^2} - 1}}
\]
It follows that the condition (\ref{RGcondition}) is equivalent to the condition that for every RT curve $B$
\be
\label{RGfinal}
\int_B T_{\mu \nu} u^\mu u^\nu ds \ge 0.
\ee
Thus, the positivity of Casini and Huerta's entanglement c-function is equivalent in holographic theories (at the classical level) to this averaged null-energy condition.\footnote{This is not equivalent to what is usually called the averaged null energy condition, which involves an average over null geodesics.} This is clearly implied by the null energy condition, but is a weaker condition, since it is possible for $T_{\mu \nu} u^\mu u^\nu \le 0 $ to be negative locally while all the integrals are positive. It may be useful to note that the condition (\ref{RGfinal}) may be expressed by saying that the ``Radon transform''\footnote{Here we mean the map from a function on a space to a function on the space of geodesic curves obtained by integrating the original function over the curve.} of $T_{\mu \nu} u^\mu u^\nu$ is everywhere non-negative.

We can give an alternative statement of the energy condition in terms of a vector field $u$ along on the curve $B$ generating a light sheet emanating from $B$. Explicitly, we can replace the condition $u \cdot \xi = 0$ with $u \cdot \partial_\lambda x_B = 0$. In terms of this null vector, the energy condition is also expressed as (\ref{RGfinal}).

\subsection{Non-monotonic scale factors}

The coordinate choice (\ref{Lorentz}) assumed the scale factor to be monotonic in the radial coordinate. In this section, we briefly consider the case where it is not. Here, we can choose coordinates
\be
\label{newcoords}
ds^2 = dr^2 + a(r)^2 (-dt^2 + dx^2) \; .
\ee
Asymptotically, $a(r)$ must be increasing, but suppose that $a'(r) < 0$ in some interval  with upper bound $r_c$, such that $a'(r_c) = 0$. Note that any such geometry violates the null energy condition $d^2/dr^2(ln(a)) \le 0$ which forbids local minima of $a$. However, we would like to understand whether such a geometry can still satisfy the constraints coming from strong subadditivity.

It is straightforward to check that $a'(r_c) = 0$ implies that $r = r_c$ is an extremal surface, so as $r_0$ approaches $r_c$, there will be a family of extremal surfaces ending on boundary intervals whose length diverges. These extremal surfaces are restricted to the region $r \ge r_c$, so their regulated length will scale with the interval size $R$ in the limit of large $R$. This is inconsistent with our assumption that the IR physics is some conformal fixed point, so it must be that beyond some $R_*$, these extremal surfaces are no longer minimal. Let $a_{1} = lim_{R \to R_*^-} a(r_0(R))$ be the minimal value of $a$ attained by this branch of extremal surfaces.

In the present coordinates, the equations for an extremal surface penetrating to some minimum radial value $r_0$ are
\[
\left({dr \over d x} \right)^2 = a^2(r) \left( {a^2(r)\over a^2(r_0)} - 1 \right) \; .
\]
Thus, we see that only when $a(r_0) = \min_{r \ge r_0} a(r)$ can an extremal surface reach the boundary. Otherwise, the previous equation would imply some negative value for $\left({dr \over d x} \right)^2$ at locations where $a(r) < a(r_0)$. Thus, the branch of extremal surfaces which become minimal for $R > R_*$ have $r_0$ greater than the value where $a(r)$ again decreases past $a(r_c)$. Let $a_2$ be the maximal value of $a$ for this $R > R_*$ branch of solutions. We see that $a_2 < a_1$.

Using the result (\ref{dSdR}) in the previous section, we have
\[
{d S \over dR} = a(r_0) \qquad \qquad c(R) = R a(r_0)
\]
so we see that non-monotonic scale factors, the entanglement c-function is discontinuous, jumping from $R_* a_1$ to $R_* a_2$ at $R = R_*$. This was emphasized previously in \cite{Myers:2012ed}.

Despite the discontinuous behavior of the RT-surfaces, the constraint from monotonicity of the c-function can still be expressed as (\ref{RGfinal}), as we can show by repeating the calculations from the previous section in the coordinates (\ref{newcoords}). In this case, the constraint applies only to the extremal surfaces with minimal area.

\section{Constraints on spacetimes dual to states of 1+1D CFTs}

In this section, we place restrictions on translation and time-translation invariant spacetimes dual to states of 1+1 dimensional holographic CFTs on Minkowski space.

\subsection{Constraints from positivity and monotonicity of relative entropy}

We start by considering constraints arising from the positivity and monotonicity of relative entropy for spacelike intervals.

For our CFT state $\Psi$, we can choose to work in a frame of reference where the stress tensor is diagonal. We consider the density matrices $\rho_I$ for a spacelike interval $I$ from $(0,0)$ to $(R_x,R_t)$. We will compare these to the density matrices $\sigma^{\beta,v}_T$ calculated from a reference state, which we take to be a boosted thermal state with temperature $\beta$ and boost parameter $v$. For these states the relative entropy $S(\rho_T || \sigma^{\beta,v}_T)$ must be positive and increase with the size of the interval,
\be
\delta^+_I S(\rho_I || \sigma^{\beta,v}_I) \ge 0
\label{REstart}
\ee
where $\delta^+_I$ represents a deformation $(R_x,R_t) \to (R_x + \delta x,R_t + \delta t)$ that increases the proper length of the interval. Note that positivity follows from this monotonicity condition since the relative entropy is zero for a vanishing interval.

According to the result (\ref{optifinal}) and the discussion in that section, the optimal relative entropy constraints will be obtained by choosing the reference state parameters $(\beta,v)$ such that the stress tensor of the boosted thermal state matches the stress-tensor of our state. This requires $v=0$ and $\beta = \beta^*$ such that the energy density of the thermal state matches that of our state. From (\ref{optifinal}) the optimal monotonicity constraint reduces simply to
\be
\label{Monotonicity}
\delta^+_I \left\{ S(\rho_I) - S(\sigma^{\beta^*}_I) \right\} \le 0
\ee
A general expression for the variation of the holographic entanglement entropy under a variation in the interval is given in appendix \ref{AppB}. The result is:
\be
\label{Sderiv}
\delta^+_I S = \delta x [A^x_0 \gamma_0] - \delta t [A^t_0 \gamma_0 \beta_0]
\ee
where $A_x^0$ and $A_t^0$ are the spatial and temporal scale factors at the deepest point $r_0$ on the extremal surface, defined for a general diagonal choice of the metric by $A_x^0 = \sqrt{g_{xx}(r_0)}$ and $A_t^0 = \sqrt{-g_{tt}(r_0)}$, and $\gamma_0 = (1 - \beta_0^2)^{-{1 \over 2}}$ with $\beta_0 = (A_t dt)/(A_x dx)$ measuring the ``tilt'' of the geodesic at the point $r_0$.

Using this result, the monotonicity constraint may be expressed as
\be
\label{mono2}
\delta x  \left\{ [A^x_0 \gamma_0]_I -  [A^x_0 \gamma_0]^{\beta^*}_I \right\} - \delta t \left\{ [A^t_0 \gamma_0 \beta_0]_I - [A^t_0 \gamma_0 \beta_0]_I^{\beta^*} \right\} \le 0
\ee
where $\Delta$ refers to difference between our state and the reference thermal state with the same stress-tensor expectation values. Here we require $\delta x > 0$ and $|\delta t| \le \delta x$, so the strongest constraint will either be for $\delta t = \delta x$ or $\delta t = - \delta x$. Thus, an equivalent statement is
\be
\label{Relent}
\Delta  [\gamma_0 (A^x_0 \pm \beta_0 A^t_0)]_I \le 0 \; ,
\ee
where $\Delta$ refers to the result for our state minus the result for the thermal state.

\subsubsection{Spatial constraint}

It is interesting to write the our constraint more explicitly for the special case of a spatial interval. We choose coordinates for which the metric takes the form
\be
\label{SFcoords}
ds^2 = {F^2(r) \over r^2} dr^2 + r^2 dx^2 - r^2 G^2(r) dt^2  \; ,
\ee
so that the radial coordinate measures the spatial scale factor. In this case, the geodesics lie on constant time slices, so $\beta_0 = 0$, $\gamma_0 = 1$, and the constraint (\ref{mono2}) gives
\be
\label{monfinal}
r_0 (R) \le r_0^{\beta^*} (R) \; ,
\ee
Thus, the monotonicity of relative entropy constraint for spatial intervals is equivalent to the statement that the minimum scale factor reached by an extremal surface in the geometry associated with $|\Psi \rangle$ is never less than the value in the thermal state geometry with the same $\langle T_{00} \rangle$.

Since $r$ is a decreasing function of $R$ according to (\ref{d2S}), the condition (\ref{monfinal}) is equivalent to
\be
\label{monfinal1}
R(r_0) \le R_\beta (r_0) \; ,
\ee
Using the coordinates (\ref{SFcoords}) and the result (\ref{defw0}), we can express this as
\be
\label{monfinal2}
\int_{1}^\infty dx {1 \over  x^2 \sqrt{x^2 - 1}} (F (r_0 x) - F_\beta(r_0 x))  \le 0 \; .
\ee
As we show in the next section, this constraint agrees asymptotically with the condition of positive energy $T_{00} \ge 0$.

More generally, we can show that the condition (\ref{monfinal2}) is implied by but does not imply the constraint of positive energy. To see this, we note that $F(\infty) = F_\beta(\infty) = 1$ and that for large $r$, $F(r) - F_\beta(r) = a r^{-n} + O(r^{-(n+1)})$ with $n \geq 3$. In our coordinates, the positive energy constraint gives $rF'(r) - F(r) + F^3(r) \geq 0$ with equality for $F_\beta(r)$ describing the thermal state. Thus,
\[
(F - F_\beta)' \ge {1 \over r}(F_\beta - F) (F_\beta^2 + F_\beta F + F^2 - 1) \; .
\]
To leading order in large $r$ this is $a (n-2) \leq 0$, so that $F(r) - F_\beta(r)$ must initially decrease below zero as we move in from $r=\infty$.
Then since $F_\beta(r) \geq 1$, $(F(r) - F_\beta(r))' \geq 0$ and $F(r) - F_\beta(r)$ must continue to decrease as $r$ decreases, ensuring that (\ref{monfinal2}) holds.

\subsubsection{Asymptotic Constraints}
\label{RE_asymptotic}

It is interesting to work out the implications of the relative entropy constraint (\ref{Relent}) on the asymptotic geometry of the spacetime. For this purpose, we choose Fefferman-Graham coordinates
\be
\label{FGexp}
ds^2 = {1 \over z^2} (dz^2 + f(z) dx^2 - g(z) dt^2) \; .
\ee
To apply the constraint (\ref{Relent}) we need an expression relating the parameters $\beta_0$, $A^x_0$, and $A^t_0$ to the parameters $(R_x,R_t)$ describing the boundary interval. Starting from the area functional
\be
\label{area2}
{\rm Area}(\tilde{B}) = \int {dz \over z} \sqrt{1 - g(z) \left({dt \over dz}\right)^2 + f(z) \left({dx \over dz}\right)^2} \; ,
\ee
we find that the surface is extremal if
\bea
{d \over dz} \left\{{ f(z) {dx \over dz} \over z \sqrt{1 - g(z) \left({dt \over dz}\right)^2 + f(z) \left({dx \over dz}\right)^2} } \right\} &=& 0 \cr
{d \over dz} \left\{{ g(z) {dt \over dz} \over z \sqrt{1 - g(z) \left({dt \over dz}\right)^2 + f(z) \left({dx \over dz}\right)^2} } \right\} &=& 0 \; .
\eea
Let $z_0$ be the maximum value of $z$ reached by the surface, and define as above
\[
\beta_0 = \sqrt{g(z_0) \over f(z_0)} {dt \over dx}(z=z_0) \; ,
\]
such that $|\beta_0| < 1 $ for a spacelike path. In terms of these parameters, we get
\bea
\left({dx \over dz}\right)^2 &=&  {z^2 f_0  \over z_0^2 f^2 } {1 \over \left[1 - {z^2 f_0  \over z_0^2 f} \right] - \beta_0^2   \left[ 1 - {z^2 g_0  \over z_0^2 g } \right]} \cr
\left({dt \over dz}\right)^2 &=&  \beta_0^2  { z^2 g_0\over z_0^2 g^2} {1 \over \left[1 - {z^2 f_0  \over z_0^2 f} \right] - \beta_0^2   \left[ 1 - {z^2 g_0  \over z_0^2 g } \right]}
\eea
where we have defined $f_0 = f(z_0)$ and $g_0 = g(z_0)$. Using these, we obtain
\bea
\label{Rdefs}
R_x &=& \int_0^{z_0} dz {z \sqrt{f_0}  \over z_0 f } {1 \over \sqrt{\left[1 - {z^2 f_0  \over z_0^2 f} \right] - \beta_0^2   \left[ 1 - {z^2 g_0  \over z_0^2 g } \right]}} \cr
R_t &=& \int_0^{z_0} dz { z  \beta_0 \sqrt{g_0}\over z_0 g } {1 \over \sqrt{\left[1 - {z^2 f_0  \over z_0^2 f} \right] - \beta_0^2   \left[ 1 - {z^2 g_0  \over z_0^2 g } \right]}}
\eea
To understand the asymptotic constraints, we can write $f$ and $g$ asymptotically as
\be
\label{fgdef}
f(z) = 1 + z^2 f_2 + z^3 f_3 + z^4 f_4 + \dots \qquad \qquad g(z) = 1 - z^2 f_2 + z^3 g_3 + z^4 g_4 + \dots \; .
\ee
where we have used tracelessness of the CFT stress tensor to conclude that
\[
[g]_{z^2} + [f]_{z^2} \propto \langle -T_{tt} + T_{xx} \rangle = 0 \; .
\]
Defining the proper length $L = \sqrt{R_x^2 - R_t^2}$ and $v = R_t/R_x$, we can use (\ref{Rdefs}) to express $L$ and $v$ as power series in $z_0$ with $\beta_0$-dependent coefficients. Inverting these, we can express $z_0$ and $\beta_0$ as power series in $L$ with $v$-dependent coefficients. Finally, we can write the expression
\[
\delta_I S = \gamma_0 (A^x_0 \pm \beta_0 A^t_0) = {1 \over \sqrt{1 - \beta_0^2}} \left({\sqrt{f(z_0)} \over z_0} + \beta_0 {\sqrt{g(z_0)} \over z_0} \right)
\]
appearing in (\ref{Relent}) as a power series in $L$ with $v$-dependent coefficients. Here we have chosen the plus sign in (\ref{Relent}) without loss of generality, since the constraint is invariant under a swap of the sign and $v -> -v$. The monotonicity constraint implies a negative difference between this expression for general $f$ and $g$ and the expression with the thermal state values
\[
f_{\beta^*} = 1 + f_2 z^2 + {1 \over 4} f_2^2 z^4 \qquad \qquad g_{\beta^*} = 1 - f_2 z^2 + {1 \over 4} f_2^2 z^4 \; .
\]
Since we are working in the limit of small $L$, the negativity implies that the leading order nonzero terms in the power series must have a negative coefficient.

In the case where $f_3$ and $g_3$ are nonzero, the leading order term is at order $L^2$, and negativity of the coefficient gives:
\[
v(3v - 2) g_3 + (2v - 3) f_3 \ge 0 \;
\]
This is required to be true for all $|v| < 1$ (corresponding to the tilt of the interval), and we find that the combination of these conditions is equivalent to
\be
\label{f3cond}
f_3 \le g_3 \qquad \qquad f_3 \le {3 \sqrt{5} - 7 \over 2} g_3 \approx -0.1459 g_3
\ee
In the case where $f_3$ and $g_3$ vanish, the constraint becomes the positivity of the $L^3$ term, which gives
\[
v(2v - 1) (g_4 - {1 \over 4}f_2^2) + (v - 2) (f_4 - {1 \over 4} f_2^2) \ge 0 \;
\]
Again, this is required to be true for all $|v| < 1$, and the combination of constraints gives
\be
\label{f4cond}
f_4  \le g_4 \qquad \qquad (f_4 - {1 \over 4}f_2^2) \le (4\sqrt{3}-7) (g_4 - {1 \over 4}f_2^2) \approx -.07178 (g_4 - {1 \over 4} f_2^2)
\ee

\subsubsection*{Comparison with standard energy conditions}

\begin{figure}
\centering
\includegraphics[width=0.8\textwidth]{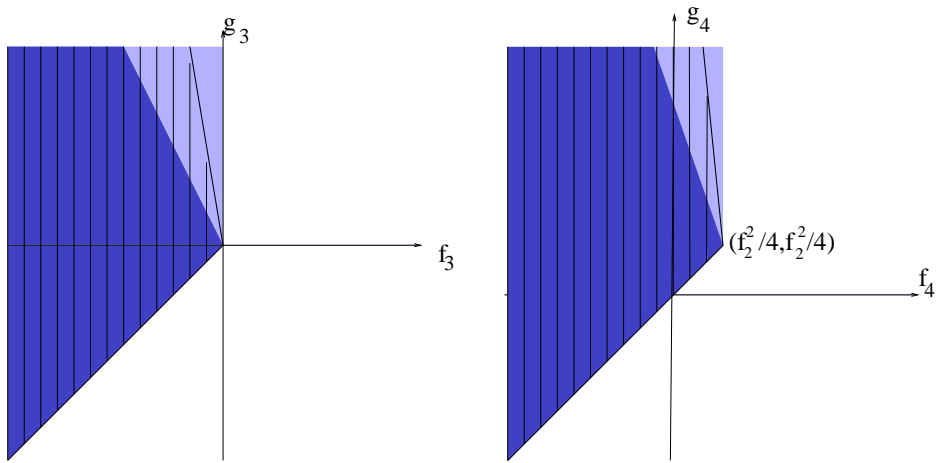}
\caption{Relative entropy constraints on coefficients in the Fefferman-Graham expansion of the metric (striped region). Constraints on the right apply only if $f_3 = g_3 = 0$. Dark blue shaded region are the constraints from the null-energy condition. Full shaded region corresponds to constraints from positivity of relative entropy, equivalent to constraints from the weak energy condition for timelike vectors with no component in the radial direction.}
\label{f3}
\end{figure}

We can compare our results to the standard weak and null energy conditions $T_{\mu \nu} u^\mu u^\nu \ge 0$ for various timelike or null vectors $u$. The non-vanishing components of the stress tensor are
\bea
\label{EEwithMatter}
T_{zz} &=& -{1 \over 2 z} {g' \over g}  -{1 \over 2 z} {f' \over f} + {1 \over 4} {f' \over f} {g' \over g} \cr
T_{tt} &=& {g \over 4 z} \left( 2 {f' \over f}  + z \left({f' \over f} \right)^2   - 2 z {f'' \over f} \right) \cr
T_{xx} &=& -{f \over 4 z} \left( 2 {g' \over g}  + z \left({g' \over g} \right)^2   - 2 z {g'' \over g} \right)
\eea
Assuming that $f_3$ and $g_3$ are nonzero, the weak energy condition applied to timelike vectors with no radial component (i.e. the non-negativity of energy for observers moving in the field theory directions) gives
\be
\label{WE3}
f_3 \le g_3 \qquad \qquad f_3 \le 0 \; ,
\ee
while including $u^\mu$ in the radial direction strengthens the conditions to
\be
\label{NEC3}
f_3 \le g_3 \qquad \qquad f_3 \le -{1 \over 2} g_3 \; .
\ee
When $f_3 = g_3 = 0$, the weak energy condition applied to timelike vectors with no radial component gives
\be
\label{WE4}
f_4 \le g_4 \qquad \qquad f_4 - {1 \over 4} f_2^2 \le 0 \; ,
\ee
while the full weak/null energy condition gives
\be
\label{NEC4}
f_4 \le g_4 \qquad \qquad f_4 - {1 \over 4} f_2^2 \le -{1 \over 3} (g_4 - {1 \over 4} g_2^2) \; .
\ee
The conditions (\ref{f3cond}) and (\ref{f4cond}) coming from monotonicity of relative entropy are intermediate between the weak/null energy condition considering only $u$ in the field theory directions and the conditions for general $u$. An interesting point is that the weaker conditions (\ref{WE3}) and (\ref{WE4}) are exactly equivalent to the conditions obtained by positivity of relative entropy (without demanding monotonicity).

\subsection{Constraints from strong subadditivity}

We now consider the constraints arising from the strong subadditivity of entanglement entropy. For a state invariant under spacetime translations, the entanglement entropy for any spacelike interval will be a single function $S(R_x,R_t)$ where $(R_x,R_t)$ represents the difference between the two endpoints. According to the discussion in section 2, the requirements of strong subadditivity in this case are implied by the minimal set of strong subadditivity constraints (\ref{SSAti}). In these formulae, we have defined $R_\pm = R_x \pm R_t$. To obtain explicit expressions for these, we can evaluate the first derivatives using the result (\ref{Sderiv}). We have
\be
\label{Sminus}
\partial_\pm S = \gamma_0 (A^x_0 \mp \beta_0 A^t_0)
\ee
where $A_t$, $A_x$, $\beta_0$, and $\gamma_0$ are defined in the previous subsection. From here, we can write the constraints (\ref{SSAti}) explicitly by taking one more derivative. For example, we have
\beas
\partial_+ \partial_- S &=& {\partial \over \partial R_+} \left[ \gamma_0 (A^x_0 + \beta_0 A^t_0) \right] \cr
&=& {\partial r_0 \over \partial R_+} {\partial \over \partial r_0} \left[ \gamma_0 (A^x_0 + \beta_0 A^t_0) \right] + {\partial \beta_0 \over \partial R_+} {\partial \over \partial \beta_0} \left[ \gamma_0 (A^x_0 + \beta_0 A^t_0) \right] \cr
&=&  {1 \over \Delta} \left\{ -{\partial R_- \over \partial \beta_0} {\partial \over \partial r_0} \left[ \gamma_0 (A^x_0 + \beta_0 A^t_0) \right] + {\partial R_- \over \partial r_0} {\partial \over \partial \beta_0} \left[ \gamma_0 (A^x_0 + \beta_0 A^t_0) \right] \right\}
\eeas
where
\[
\Delta = \det \left( \ba{cc} {\partial R_- \over \partial r_0} & {\partial R_- \over \partial \beta_0} \cr {\partial R_+ \over \partial r_0} & {\partial R_+ \over \partial \beta_0} \ea \right) \; .
\]
The strong subadditivity constraint is then that $\partial_+ \partial_- S \le 0$. Here, the determinant $\Delta$ is positive for geometries in some neighborhood of pure AdS (and possibly more generally); in this case, the constraint simplifies to the statement that the expression in curly brackets is non-positive.

We can write an explicit expressions for $R_-$ and $R_+$ using the steps leading to (\ref{Rdefs}). We find
\be
\label{Rdefs1}
R_\pm = \int_\gamma ds \gamma_0 \left\{{A_0^x \over (A^x)^2} \pm \beta_0 {A_0^t \over (A^t)^2} \right\}
\ee
where the integral is along the extremal surface, with length element
\[
ds =  {dr \sqrt{g_{rr}}  \over \gamma_0 \sqrt{\left[1 - {(A^x(r_0))^2  \over (A^x(r))^2} \right] - \beta_0^2   \left[ 1 - {(A^t(r_0))^2 \over (A^t(r))^2} \right]}} \; .
\]
From this, the constraint $\partial_+ \partial_- S \le 0$ for each spacelike interval $I$ can be expressed as an integral over the extremal curve $\gamma$ ending on $I$. It is natural to expect that the result can be expressed in a covariant form similar to (\ref{RGfinal}), but we leave this for future work.

\subsubsection*{Asymptotic constraints}

Using the tools from section \ref{RE_asymptotic}, it is straightforward to work out the constraints on the asymptotic geometry implied by the strong subadditivity constraint $\partial_+ \partial_- S \le 0$. Note that the conditions $\partial_+ \partial_+ S \le 0$ and $\partial_- \partial_- S \le 0$ are always satisfied asymptotically.

We work again in the Fefferman-Graham expansion (\ref{FGexp}) with metric functions expanded as (\ref{fgdef}). We can write the expression (\ref{Sminus}) as a power series in the proper length $L$ of the interval, with coefficients depending on the ratio $\beta = R_t/R_x$ and the coefficients appearing in (\ref{fgdef}). Acting with
\beas
\partial_+ &=& {\partial L \over \partial R_+} \partial_L + {\partial v \over \partial R_+} \partial_v \;  \cr
&=& {1 \over 2} \sqrt{1 - v \over 1 + v} \left\{ \partial_L + (1 - v^2) {1 \over L} \partial_v \right\}
\eeas
gives a power series for $\partial_+ \partial_- S$, and the strong subadditivity constraint implies that the leading non-zero coefficient must be negative.

In the case where $f_3$ and $g_3$ are nonzero, the leading order term is at order $L$, and negativity of the coefficient gives:
\[
(2- 7 v^2) g_3 \le - (7 - 2 v^2) f_3  \;
\]
This is required to be true for all $|v| < 1$ (corresponding to the tilt of the interval), and we find that the combination of these conditions is equivalent to
\be
\label{f3condSS}
f_3 \le g_3 \qquad \qquad f_3 \le -{2 \over 7} g_3
\ee
In the case where $f_3$ and $g_3$ vanish, the constraint becomes the negativity of the $L^2$ term, which gives
\be
(1- 7 v^2) (g_4 - {1 \over 4} f_2^2) \le - (7 - v^2) (f_4 - {1 \over 4} f_2^2) \ge 0 \;
\ee
Again, this is required to be true for all $|v| < 1$, and the combination of constraints gives
\be
\label{f4condSS}
f_4  \le g_4 \qquad \qquad (f_4 - {1 \over 4}f_2^2) \le -{1 \over 7} (g_4 - {1 \over 4}f_2^2)
\ee
These constraints take a similar form to the constraints (\ref{f3cond}) and (\ref{f4cond}) from monotonicity of relative entropy, but are slightly stronger. However, they are still weaker than the constraints (\ref{NEC3}) and (\ref{NEC4}) arising from the null energy condition.

\section{Constraints on spherically-symmetric asymptotically AdS spacetimes}

In this section, we point out a simple constraint on the geometries of static, spherically symmetric asymptotically $AdS_{d+2}$ spacetimes. This would apply for example to spherically symmetric ``stars'' made of any allowable type of matter in a theory of gravity whose classical limit is Einstein gravity coupled to matter.

For these spacetimes, the dual state is an excited state of the dual CFT on a sphere with a homogeneous stress tensor. If the mass of the spacetime (relative to empty AdS) is $M$, the field theory energy is $M \ell$ (taking the sphere radius equal to one for the CFT), so we can say that the energy density expectation value for this state relative to the vacuum state is
\be
\label{spherestress}
\Delta \langle T_{00} \rangle = {M \ell \over \Omega_d} \; ,
\ee
where $\Omega_d$ is the volume of a $d$-sphere.

Now, consider a ball-shaped region $B_\theta$ of angular radius $\theta_0$ on the sphere. For this region, the relative entropy for our state with respect to the vacuum state is
\beas
S_{B_\theta}(\rho || 0 ) &=& \Delta \langle H_{mod} \rangle - \Delta S \cr
&=& 2 \pi \int_B d \Omega_d  {\cos(\theta) - \cos(\theta_0) \over \sin(\theta_0)} \Delta \langle T_{00} \rangle - \Delta S
\eeas
where we have used the expression (\ref{Hmodsphere}) for the modular Hamiltonian.

Since the stress tensor (\ref{spherestress}) is constant on the sphere, we can perform the integral explicitly to obtain
\beas
S_{B_\theta}(\rho || 0 ) &=& - \Delta S + {2 \pi  M \ell \Omega_{d-1} \over \Omega_d} I_d(\theta_0)
\eeas
where
\[
I_d(\theta_0) = \int_0^{\theta_0} d \theta  \sin(\theta)^{d-1} {\cos(\theta) - \cos(\theta_0) \over \sin(\theta_0)} = {(\sin \theta_0)^{d-1} \over d} \left[1 -  {}_2 F_1\left({1 \over 2}, {d \over 2} ; {d \over 2} + 1 ;  \sin^2 \theta_0 \right)\cos \theta_0 \right]\; .
\]
Then, using the Ryu-Takayanagi formula, the positivity of relative entropy gives the constraint
\[
\Delta {\rm Area}(\theta_0) \le 8 \sqrt{\pi} G_N M \ell I_d(\theta_0) {\Gamma \left({d \over 2}+{1 \over 2} \right) \over \Gamma \left({d \over 2} \right)}\; .
\]
where $\Delta {\rm Area}$ is the area of the bulk extremal surface with boundary $\delta B_\theta$.

For the special case of a hemisphere ($\theta_0 = \pi/2$), we have that
\[
\Delta {\rm Area}(\pi/2) \le 8 \sqrt{\pi} G_N  M \ell {\Gamma \left({d \over 2}+{1 \over 2} \right) \over d \Gamma \left({d \over 2} \right)} \; .
\]
which reduces for 3+1 dimensional gravity to
\[
\Delta A \le 2 \pi G_N  M \ell_{AdS} \; .
\]
Typically, the minimal area extremal surface bounded by an equator on the sphere will be the surface bisecting the spacetime symmetrically, so this constraint bounds the change in area for this bisecting surface by the mass contained in the spacetime.\footnote{In some cases, however, there may exist more than one extremal surface bounded by an equator, and in this case, the minimal area surface may not be the symmetrical one.} Roughly, the constraint places a bound on how much a certain amount of total energy in the spacetime can curve the spacetime.

\section{Discussion}

In this paper, we have explored constraints from entanglement inequalities on highly symmetric spacetimes. It will be interesting to see how these results generalize to less symmetric cases. In our analysis, we have used only the classical term in the Ryu-Takayanagi formula, so our constraints apply to gravitational theories in the classical limit. It would be interesting to understand  how the constraints are corrected when the contribution of bulk quantum fields are taken into account. This should be possible using the quantum-corrected holographic entanglement entropy formula proposed by \cite{Faulkner:2013ana}.

\subsection{Constraints on entanglement structure from geometry}

Before concluding, we offer a few remarks on the orthogonal research direction of understanding which entanglement structures are consistent with the existence of a geometrical dual spacetime. In the language of figure 1, we would like to precisely characterize the image of ${\cal G}$ in ${\cal S}$ (or in (${\cal S}_{phys}$). Here, we make a few qualitative observations that hopefully illuminate how severe these constraints are.

Consider a general asymptotically $AdS_{d+2}$ spacetime. In a Fefferman-Graham description of the metric,
\[
ds^2 = {1 \over z^2} \left[ dz^2 + \Gamma_{\mu \nu}(z,x) dx^\mu dx^\nu \right]
\]
the information about the geometry is contained in the functions $\Gamma_{\mu \nu}(z,x)$ of $(d+1)$ variables.

A set of entanglement entropies that includes a similar amount of information as one of these functions is the set $\{S(R,x)\}$ for ball-shaped regions with any radius $R$ centered at any point $x$. At least close to the boundary (where the geometry is similar to $AdS$), we expect that there is a one-to-one correspondence between pairs $(R,x)$ and bulk points $(z,x_{bulk})$, obtained by choosing the point on the RT surface with the largest value of $z$. For pure AdS, we have simply $(z,x_{bulk}) = (R,x)$. Thus, given the entanglement entropies for ball-shaped regions in one spatial slice, it is plausible that we can reconstruct some combination of the metric functions $\Gamma_{\mu \nu}(z,x)$. The other combinations are related by Lorentz-transformations, so it is further plausible that we can reconstruct the remaining functions (in some neighborhood of the boundary) by considering entanglement entropies for ball-shaped regions in other Lorentz frames.

Assuming this reconstruction is possible, we now have enough information (the full geometry in a neighborhood of the boundary) to calculate entanglement entropies for regions of any other shape. Thus, it is plausible that {\it for a quantum state with gravity dual, the entanglement entropies for regions of arbitrary shape (assuming they are not too large) are completely determined from the entanglement entropies for ball-shaped regions} (in the various frames of reference). Furthermore, they are determined in a very specific way, via construction of a dual geometry and calculation of extremal surface areas. A natural question is then to understand which field theory Hamiltonians can give rise to low-energy states with this entanglement structure, and/or why the known examples of holographic CFTs have this property.

\section*{Acknowledgements}

We thank Raphael Bousso, Laurent Chaurette, Thomas Hartman, Juan Maldacena, Hirosi Ooguri, and Brian Swingle for helpful discussions. The research of NL, CR and MVR is supported in part by the Natural Sciences and Engineering Research Council of Canada. MVR and PSG are supported in part by FQXi and FRQNT, respectively.

\appendix

\section{Modular Hamiltonian for an interval in a boosted thermal state of a 1+1D CFT}\label{AppC}

In this appendix, we derive the modular Hamiltonian for a spatial interval $[-R,R]$ in the boosted thermal state. To do this, we start by considering the domain of dependence $D_1$ of the interval $[-r,r]$ for the vacuum state in Minkowski space with coordinates $(t',x')$. For this interval, the modular Hamiltonian is quantum operator associated with the conformal isometry generated by
\[
H_1 = {\pi \over r} ((r^2 - (t')^2 - (x')^2) \partial_{t'} - 2 t' x' \partial_{x'}) \; .
\]
We can now apply a boost
\[
x' = \gamma(x - vt) \qquad \qquad t' = \gamma ( t - v x) \; .
\]
In this case, the region $D_1$ maps to the domain of dependence $D_2$ of the interval from $-(r_t, r_x)$ to $(r_t, r_x)$, where $r^2 = r_x^2 - r_t^2$ and $v = r_t/r_x$. In this case, the generator $H_1$ maps to
\[
H_2 = {\pi \over r_x^2 - r_t^2} \left[(r_x(r_x^2 - r_t^2) + 2 t x r_t - r_x (t^2 + x^2))\partial_t + (r_t(r_x^2 - r_t^2)- 2 t x r_x + r_t (x^2 + t^2))\partial_x \right]
\]
Next, we perform a transformation for which the causal development of the interval $[-1,1]$ maps to the full Minkowski space (with coordinates $(u,\tau)$), such that the resulting state is the thermal state on Minkowski space dual to the planar BTZ geometry with horizon at $z=z_0$.
The appropriate transformation (which can be obtained by finding the coordinate transformation that maps the bulk region associated with the domain of dependence of $[-1,1]$ to the planar BTZ black hole) is
\bea
t &=& {\sinh(2 \tau/z_0) \over \cosh(2 u/z_0) + \cosh(2 \tau/z_0)} \cr
x &=& {\sinh(2 u/z_0) \over \cosh(2 u/z_0) + \cosh(2 \tau/z_0)} \; .
\eea
After the map, the region $D_2$ maps to the domain of dependence $D_3$ of the interval from $-(R_t, R_u)$ to $(R_t, R_u)$, where
\bea
r_t &=& {\sinh(2 R_t/z_0) \over \cosh(2 R_u/z_0) + \cosh(2 R_t/z_0)} \cr
r_x &=& {\sinh(2 R_u/z_0) \over \cosh(2 R_u/z_0) + \cosh(2 R_t/z_0)} \; .
\eea
The generator $H_2$ maps to
\bea
H_3 &=& {\pi z_0 \over  C_u^2 - C_t^2}\left[\left\{C_u S_u + C_u S_t \sinh(2 u/z_0) \sinh( 2 \tau/z_0) -  C_t S_u \cosh(2  u/z_0) \cosh( 2  \tau /z_0 \right\}\partial_\tau \right. \cr
&& \qquad \qquad \left. \left\{-C_t S_t + C_u S_t \cosh(2 u /z_0) \cosh( 2 \tau /z_0) -  C_t S_u \sinh(2 u/z_0) \sinh( 2  \tau /z_0 \right\}\partial_u \right] \cr
\eea
where
\[
C_u = \cosh(2 R_u /z_0) \qquad S_u = \sinh(2 R_u /z_0) \; .
\]
Finally, we can perform one further Lorentz transformation
\[
u = \gamma(u' + v \tau') \qquad \qquad \tau = \gamma ( \tau' + v u') \; .
\]
with velocity $v = R_t/R_x$, such that the region $D_3$ is mapped to the domain of dependence of the interval $[-R,R]$, where $R^2 = R_x^2 - R_t^2$. In terms of $v, z_0,$ and $R$, we find that the generator $H_3$ restricted to $\tau' = 0$ gives
\bea
H_4 = {\pi \gamma z_0 \over  C_u^2 - C_t^2} \left\{ -\partial_{\tau'} \right. && \left( \cosh(\gamma v U) \cosh(\gamma U) (C_u S_t v + S_u C_t)  \right. \cr
&& - \sinh(\gamma v U) \sinh(\gamma U) (S_u C_t v + S_t C_u) \cr
 &&\left.-(S_t C_t v + C_u S_u) \right) \cr
+\partial_{u'} && \left(  \cosh(\gamma v U) \cosh(\gamma U) (C_u S_t  + S_u C_t v)  \right. \cr
&& - \sinh(\gamma v U) \sinh(\gamma U) (S_u C_t  + S_t C_u v) \cr
 &&\left. \left. -(S_t C_t + C_u S_u v) \right) \right\}
\eea
where we define $U = 2 u' /z_0$ and
\[
C_t = \cosh(2 R \gamma v /z_0) \qquad C_u = \cosh(2 R \gamma /z_0) \qquad S_t = \sinh(2 R \gamma v /z_0) \qquad S_u = \sinh(2 R \gamma /z_0) \; .
\]
The modular Hamiltonian is obtained by making the replacements $\partial_{\tau'} \to T_{\tau' \tau'}$ and $\partial_{u'} \to T_{\tau' u'}$ and integrating over $[-R,R]$.

\section{Variation in geodesic length under endpoint variation}\label{AppB}

In this section, we derive a formula for the variation of the entanglement entropy of a boosted interval for some translation and time-translation invariant state in a holographic 1+1 dimensional field theory under a general variation in the endpoint of the interval.\footnote{It is interesting to note that techniques similar to those in this section were used in \cite{Headrick:2014eia} to show a relation between differential entropy and the lengths of bulk curves.} We assume that the field theory lives on Minkowski space with coordinates $(x,t)$.

The dual spacetime will be a 2+1 dimensional spacetime with translational isometries in one spatial direction and one time direction, associated with Killing vectors $\xi_t^\mu$ and $\xi_x^\mu$. We assume that the spacetime has a conformal boundary, with a Minkowski space boundary geometry $ds^2 = -dt^2 + dx^2$ such that the Killing vectors $\xi_t^\mu$ and $\xi_x^\mu$ become $\partial_t = (1,0)$ and $\partial_x = (0,1)$ at the boundary. Consider a spatial geodesic with endpoints on the boundary at points 0 and $R(\gamma , \gamma v)$, where $v < 1$, $\gamma = (1 - v^2)^{-1}$. We would like to determine the variation in length of the geodesic under a variation in the proper length $R$ of the boundary interval.

The geodesic is an extremum of the action
\be
\label{genac}
S = \int_i^f d \lambda \sqrt{g_{\mu \nu} {d x^\mu \over d \lambda} {d x^\nu \over d \lambda}} \; .
\ee
In general, the variation of an action $S = \int d \lambda {\cal L}(q_n, \dot{q}_n)$ evaluated for an on-shell configuration under a variation of the boundary conditions (assuming the range of integration remains the same) is given by
\[
\delta S = \left[p_n \delta q_n \right]_i^f \; ,
\]
where $q_n$ are the coordinates and $p_n = \partial {\cal L} / \partial q_n$ are the conjugate momenta. This follows immediately since the variation of the action gives a total derivative when the Euler-Lagrange equations are satisfied. Consider a general variation of the endpoints
\[
\delta x^\mu_f = \delta x \xi^\mu_x + \delta t \xi^\mu_t\; .
\]
Since the conjugate momentum to $x^\mu$ is
\[
p_\mu = {\partial {\cal L} \over \partial x^{\mu}} = {g_{\mu \nu} {d x^\nu \over d \lambda} \over \sqrt{g_{\mu \nu} {d x^\mu \over d \lambda} {d x^\nu \over d \lambda}}} \; .
\]
we have
\be
\delta S = \delta x \xi^\mu_x p_\mu + \delta t \xi^\mu_t p_\mu  \; .
\label{delS1}
\ee
Now, for a Killing vector $\xi^\mu$, the action (\ref{genac}) is invariant under $x^\mu \to x^\mu + \xi^\mu$. The corresponding conserved quantity is exactly $\xi^\mu p_\mu$. Thus, the right hand side of (\ref{delS1}) can be evaluated at any point on the trajectory. We choose to evaluate it at the midpoint of the geodesic, where $\partial_\lambda x^\mu$ is a linear combination of
$\xi^\mu_t$ and $\xi^\mu_x$ (i.e. with no component in the radial direction). In this case,
\[
\partial_\lambda x^\mu = \xi^\mu_t {\xi_t \cdot \partial_\lambda x \over \xi_t \cdot \xi_t} + \xi^\mu_x {\xi_x \cdot \partial_\lambda x \over \xi_x \cdot \xi_x} \; ,
\]
so we find that our expression (\ref{delS1}) becomes
\be
\delta S  =  \delta x [\gamma_0 A^x_0 ]  + \delta t [\gamma_0 \beta_0 A^t_0]   \; .
\label{delS2}
\ee
where we have defined
\beas
A^x_0 &=& \sqrt{\xi_x \cdot \xi_x} \cr
A^t_0 &=& \sqrt{-\xi_t \cdot \xi_t} \cr
\beta_0 &=& {A^x_0 \over A^t_0} {\xi_t \cdot \partial_\lambda x \over \xi_x \cdot \partial_\lambda x} \cr
\gamma_0 &=& {1 \over \sqrt{1 - \beta_0^2}}\; ,
\eeas
which measures the ``tilt'' of the geodesic at the midpoint.

In the special case of a spatial interval, we will have $\xi_t \cdot \partial_\lambda x = 0$ everywhere, so
\be
{\delta S \over \delta R} = \sqrt{\xi_x^2} = \sqrt{g_{\mu \nu} \xi_x^\mu \xi_x^\nu}\; .
\ee
Thus, the variation of the entanglement entropy with respect to the size of a spatial interval gives exactly the spatial scale factor.

\bibliographystyle{JHEP}

\bibliography{relative_constraint}

\end{document}